\documentclass[runningheads]{llncs}
\usepackage{graphicx}
\usepackage{color}

\usepackage[bookmarks, colorlinks, breaklinks]{hyperref}

\begin{document}
\title{
  Non(c)esuch Card-Level Comparison Risk-Limiting Audits}
  \titlerunning{Non(c)esuch RLAs}

\author{
Philip B. Stark\inst{1}\orcidID{0000-0002-3771-9604} 
}
\authorrunning{P.B. Stark}
%
\institute{University of California, Berkeley, CA USA
\email{stark@stat.berkeley.edu}}
\maketitle              
\begin{abstract}
    Risk-limiting audits (RLAs) guarantee a high probability of correcting incorrect reported electoral
    outcomes before the outcomes are certified.
    The most efficient are \emph{card-level comparison audits} (CLCAs), which compare
    the voting system's
    interpretation of randomly selected individual ballot cards (\emph{cast-vote records}, CVRs)
    from a trustworthy paper trail to a human interpretation
    of the same cards.
    CLCAs have logistical and privacy hurdles: Individual randomly selected cards must be retrieved for manual inspection;
    the voting system must export CVRs; and the CVRs must be linked to the corresponding physical cards, to compare the two.
    In practice, such links have been made by keeping cards in the order in which they are scanned or by printing
    serial numbers on cards as they are scanned. 
    Both methods may compromise voter privacy.
    Cards selected for audit have been retrieved by manually counting into stacks or by looking for cards with
    particular serial numbers.
    The methods are time-consuming; the first is also error-prone.
    Connecting CVRs to cards using a unique pseudo-random number (``cryptographic nonce'') printed
    on each card after the voter last sees it could reduce privacy risks,
    but retrieving the card imprinted with a particular random number may be harder than counting
    into a stack or finding the card with a given serial number.
    And what if the system does not in fact print a unique number on each ballot or does not
    accurately report the numbers it printed?
    This paper presents a method for conducting CLCAs that maintains the risk limit
    even if the system does not print a genuine nonce on each ballot or misreports the identifiers it used.
    The method also allows untrusted technology to be used to retrieve the cards selected for audit---automation
    that may drastically reduce audit workload, whether cards are imprinted with serial numbers or putative nonces.
    The method limits the risk rigorously, even if the imprinting or retrieval technology misbehaves.
    If the imprinting and retrieval systems behave properly, this protection does not increase the number of
    cards the RLA has to inspect to confirm or correct the outcome. 
\end{abstract}


\section{Introduction: Efficient Risk-Limiting Audits}
A risk-limiting audit (RLA) is any procedure that has a known minimum chance
of correcting the reported election outcome if the reported outcome is wrong,
and never changes a correct outcome.
The largest chance that the audit fails to correct a wrong outcome
is the \emph{risk limit}.
(The \emph{outcome} is the political outcome---who or what won---not the exact vote tally.)

RLAs require a trustworthy record of the validly cast votes, which generally entails
conducting the election using hand-marked paper ballots \cite{appelEtal20,appelStark20,starkXie22}
kept demonstrably secure throughout the election, canvass, and audit.
Eligibility audits and compliance audits are crucial to check whether the paper trail is trustworthy \cite{appelStark20}.
There are RLA methods based on sampling individual ballots and clusters of ballots,
with or without replacement, with or without stratification, with or without
sampling weights, and for Bernoulli sampling 
\cite{stark08a,stark09b,higginsEtal11,ottoboniEtal18,ottoboniEtal20,stark20,stark22}.

There are many ways to use data from randomly selected, manually interpreted ballot cards to
conduct an RLA.
\emph{Polling audits} involve manually reading votes from ballot cards, but do not
require any information from the voting system other than the reported winner(s).
\emph{Comparison audits} involve comparing votes read manually from ballot cards to 
to how the voting system interpreted the same cards.
There are also \emph{hybrid} audits, which combine polling and comparison \cite{ottoboniEtal18}.

The most efficient risk-limiting methods (those that manually examine the fewest
ballot cards when reported outcomes are correct) are \emph{card-level comparison audits} (CLCAs),
which compare how the voting equipment interpreted individual, randomly selected ballot cards 
(\emph{cast-vote records}, CVRs) to how
humans interpret the same cards (\emph{manual-vote records}, MVRs).
CLCAs require the voting system to commit to CVRs before the audit starts,
in such a way that individual CVRs can be matched to individual ballot cards. 
Constructing that matching can be challenging.
Strategies used in various jurisdictions include imprinting identifiers on ballot
cards after the cards have been disassociated from voters but before or while 
the cards are scanned, or simply trying to keep the ballot cards in the same
order in which they were scanned.
Counties in the State of Colorado use both of those methods.

Imprinting ballot cards or keeping track of scan order may work when
ballots are scanned centrally (central-count optical scan, CCOS) or in vote centers.
But when ballots are scanned in precincts (precinct-count optical scan, PCOS),
imprinting serial numbers on ballots or tracking scan order may compromise the
anonymity of votes: for instance, a careful observer might be able to tell that a given voter
was the 17th voter to cast a ballot using a particular scanner.
While it is tacitly assumed that the act of casting ballots randomizes their order,
I am unaware of any study of the order in which
ballot cards land in the ballot box compared to the order in which they are cast.
Unless the cross section of a ballot box is at least twice the size of a ballot in at least one dimension, 
it seems physically implausible that ballots would
be shuffled well by the act of casting.
Thus, unless the cast physical ballots are mechanically shuffled before anyone has the opportunity
to inspect them, it is plausible that an insider can exploit cast card order to compromise vote anonymity,
whether ballots are imprinted with identifiers or not.

To take full advantage of the efficiency of CLCAs, some jurisdictions
have re-scanned PCOS ballots centrally and based the audit on that rescan.\footnote{%
E.g., Rhode Island \url{https://www.brennancenter.org/sites/default/files/2019-09/Report-RI-Design-FINAL-WEB4.pdf}.
}
This entails considerable duplication of effort; moreover, it checks the second tabulation, not the official, original tabulation.
Other jurisdictions\footnote{%
E.g., Michigan 
\url{https://www.brennancenter.org/sites/default/files/2019-11/2019\_011\_RLA\_Analysis\_FINAL\_0.pdf}.
}
have piloted \emph{hybrid} audits that use stratified sampling: card-level comparison in one stratum and ballot polling 
in another, combining the results using union-intersection tests (SUITE, \cite{ottoboniEtal18}).
CVRs from a 
a random sample of cards tabulated using CCOS are compared to a manual reading of the votes 
from those cards, and an independent sample of cards tabulated using PCOS
are manually interpreted but not compared to their machine interpretations.
(An approach to stratified and hybrid audits that improves on SUITE was introduced by 
\cite{stark22}; 
its performance was explored by \cite{spertusStark22}.)

A third option is to use card-level comparison for CCOS and batch-level comparison for
PCOS \cite{ottoboniEtal18}, i.e., a batch-level comparison audit with batches that range in 
size from a single ballot card to all ballot cards cast in a precinct.
Only the first of these approaches yields the audit-day efficiency of card-level 
comparison audits,
but at the cost of rescanning all PCOS ballots.
A fourth approach is ``lazy'' card-level comparison audits  \cite{harrisonEtal22}, which 
also involve an original batch tabulation for the official election result;
that tabulation does not produce CVRs linked to individual cards.
The lazy audit involves a re-scan to generate CVRs linked to ballots---but 
only for batches from which a ballot card is selected for audit, potentially
saving work if
the sample size is small enough that only a fraction of batches need to be re-scanned.
\cite{harrisonEtal22} derive such a ``lazy'' audit of
single-winner plurality contests, and find that
under some conditions (relating to the sizes of precincts, the fraction of CCOS versus PCOS votes, margins, etc.), 
``lazy'' audits can be more efficient than ``hybrid'' audits \cite{ottoboniEtal18}.

If the mapping from CVRs to cards is based on the scan order, retrieving particular ballot cards
for the audit involves counting into stacks of ballot cards that, in many jurisdictions, contain hundreds of cards.
This manual process is time-consuming and error prone.
If the mapping is based on imprinting serial numbers, retrieving individual cards is generally easier, unless the 
card order changes, e.g., because a batch of ballots is dropped.
With the exception of \cite{harrisonEtal22}, methods that link CVRs to cards by printing serial numbers on ballots have not 
contemplated whether the audit still rigorously limits the risk if something goes wrong with the imprinting, 
for instance, if the imprinting omits or re-uses identifiers.

This paper shows that imprinting PCOS ballot cards with
unpredictable identifiers as the cards are scanned could make efficient
card-level
comparison audits of PCOS systems possible by allowing
ballot cards to be associated with their CVRs while obfuscating the link between any ballot
card and the particular voter who cast that card---and without trusting the technology used to imprint the identifiers.
It also shows that untrusted technology (including untrusted auditors!) can be used to retrieve ballot cards.

The method presented here has some features in common with the method of
\cite{harrisonEtal22}, which also contemplates untrusted retrieval of cards, with the following differences:
\begin{itemize}
   \item The method does not involve re-scanning, only an initial scan.
   \item The method does not require a separate ``commitment'' to batch subtotals, only to the CVRs.
   \item The proof that the method rigorously limits the risk is different and more general.
   \item The method works for every social choice function for which there is a known RLA method,
including plurality, multi-winner plurality, supermajority, Borda count, approval voting,
all scoring rules, STAR-Voting, proportional representation methods such as D'Hondt and Hamilton, and instant-runoff voting (IRV).
\end{itemize}

\paragraph*{Nonces, IDs, and privacy.} 
A cryptographic \emph{nonce}\footnote{%
See \url{https://csrc.nist.gov/glossary/term/nonce},
last visited 4 July 2022)}
(number once) is a number guaranteed to be unique (in some universe
of numbers).
Nonces are typically generated randomly or pseudo-randomly.
If we could trust software and hardware to print a genuine nonce on each ballot card, the nonces could serve as IDs: 
each card would have a unique ID 
linking 
it to its CVR, but the randomness would ensure that there was no information in the ID
that could compromise vote privacy.

However, such software should not be trusted.
A faulty or malicious implementation---or knowledge of the seed in the
pseudo-random number generator (PRNG) used to generate nonces---could compromise vote
anonymity by allowing the order of casting to be recovered.\footnote{%
The DVSOrder vulnerability of some Dominion systems, published in October 2022, illustrates this problem. 
See \url{https://dvsorder.org}.
(Last visited 27 January 2023.)
}
This paper addresses different issues:
if the system cannot be trusted to imprint ballot cards with the IDs it claimed to use, 
can an RLA still use the reported IDs to link cards to CVRs? 
How?
Since finding a card imprinted with a particular nonce is hard, can an RLA use untrusted technology
to retrieve cards? How?

\paragraph*{The crux of the matter.}
What if the imprinter prints the same ID on more than one ballot, misreports the IDs it printed, or fails to
imprint an ID on one or more ballot cards?
Could that undermine an RLA?

To see why that is a problem, consider a two-candidate plurality contest, Alice v. Bob.
An attacker with control of the imprinting wants Bob to win.
Suppose 3~ballot cards were cast, two showing votes for Alice and one for Bob:
Alice really won.
The attacker creates 3~cast vote records: the CVR with ID 17 has a vote for Alice; CVRs with the
IDs 91 and 202 have a vote for Bob.
According to the CVRs, Bob won.
The attacker prints the number 17 on the two cards with votes for Alice and the number 91
on the card with a vote for Bob.

Auditors select a ballot card at random, for instance using the $K$-cut method of
\cite{sridharRivest20}.
They read the ID imprinted on the card.
Then they look up
the corresponding CVR.

That CVR will match the human reading of the ballot perfectly, no matter which
ballot was selected: there is zero chance of discovering the attack.
If additional ballot cards are drawn at random, there will be a nonzero probability of 
noticing that two ballots had been assigned the same ID, but as the
number of ballots grows, the chance of detecting a duplicate in a given number of
draws shrinks.
Hence, sampling physical ballot cards---rather than sampling CVRs---requires verifying that 
the identifiers printed on the ballot cards are unique.

The uniqueness of the imprinted IDs might be checked by scanning the imprinted IDs 
and processing the images, 
but that entails trusting
hardware and software to identify and report mismatches.
Hence, auditing by sampling physical ballot cards requires manually inspecting every cast card 
to compile a list of IDs and checking that no ID was repeated.\footnote{%
  However, it does not require checking the list against the system's reported list of IDs:
  suppose that the ID on a card does not match any ID in the list of CVRs.
  Treating the card as if its CVR were as unfavorable as possible to every outcome
  (an ``evil zombie'' in the terminology of \cite{banuelosStark12,stark20}), e.g., as if it showed
  a valid vote for every loser in a plurality contest, ensures that the audit will not stop
  sooner than it would have stopped if the list of IDs had been accurate.
}  

As mentioned above, checking that every card is imprinted with a unique ID can ensure 
that an RLA based on sampling ballot cards genuinely limits the risk, 
but does not ensure
that the purported nonces cannot be ``reverse engineered'' to infer the order in which the cards
were cast.
That is, the check can ensure that the reported winner(s) really won, but cannot ensure
the anonymity of the votes.

In the example above, non-uniqueness of imprinted IDs is not a problem if we sample 
CVRs and retrieve the corresponding
cards, rather than sample cards and look up the corresponding CVRs,
because there would be some chance that we select CVR~202 and discover
that no ballot card has that ID.
If we treat the ``missing'' card in the least favorable way for the audit (i.e., as a vote for Alice, the reported loser), the audit would still limit the risk to its nominal level. 
But the logistics of looking through a large pile of paper for a card imprinted 
with a particular identifier (pseudo-random nonce or serial number) is labor-intensive and error-prone.

Can we use technology to search for a card with the selected ID (nonce or serial), 
without having to trust that technology?
For instance, what if the retrieval system returns a card with a different ID, or no card at all---for instance, if returning the
correct card would have cast doubt on the outcome? 
What if more than one card is labeled with the same ID, and the retrieval system picks the card to return adversarially?
This paper provides a method to use untrusted retrieval systems without compromising the risk limit.

\section{Assumptions}
We assume that the paper trail---the collection of ballot cards---consists of every validly cast card in the contest(s) under audit, and that those cards
accurately reflect the voters' selections.
As mentioned previously, this generally requires that
cards were hand-marked by the voters and kept demonstrably secure, that the integrity of the paper trail was
verified by a compliance audit, and that eligibility 
determinations have also been checked.

We assume that there is a trustworthy upper bound on the total number of validly cast cards and on the number of validly cast cards that contain each
contest under audit.
These might be derived from a combination of voter registration records, voter participation records (including pollbooks), and physical ballot accounting.

We consider the scanning and tabulation system, the imprinting system, and the card-retrieval system to be untrusted.
How might this overall system misbehave? 

\begin{enumerate}
   \item The number of CVRs that contain the contest might not be equal to the number of cards that contain the contest.
   \item The CVR list might misrepresent the votes on one or more cards.
   \item The imprinting might omit or repeat one or more of the IDs in the CVR list, print one or more 
      IDs that are not in the CVR list, or fail to imprint an ID on one or more cards.
   \item The software could return a ballot card with 
      a different ID than the ID requested, return
      a card with no ID, or fail to return any 
      card whatsoever.
\end{enumerate}

We must demonstrate an audit procedure that has a 
guaranteed minimum chance $1-\alpha$ of proceeding to a full
hand tabulation of the cards
if the outcome according to the CVRs differs from
the outcome according to the votes on the trustworthy collection of ballot cards.
That procedure is then an RLA with risk limit $\alpha$.

\subsection{SHANGRLA assorters}

SHANGRLA \cite{stark20} reduces RLAs to statistical tests of whether the averages of
a collection of finite lists of nonnegative, bounded numbers are all greater than
$1/2$.
Each list results from applying an ``assorter'' $A$ to the votes in a contest
for each validly cast card that contains the contest (for audits that use card style data, as described in \cite{glazerEtal21})
or for all validly cast cards.
The number of assorters involved in auditing a contest depends on the social
choice function for that contest, among other things.
For instance, for a $K$-winner plurality contest in which there are $C$ candidates,
there are $K(C-K)$ assorters.
The smallest value an assorter $A$ can assign to any card or CVR is 0.
The largest value, $u$, depends on the social choice function and on whether the
audit is a comparison audit or polling audit, among other things
\cite{stark20}.

Assorters can be constructed to audit every social choice function for which an RLA method is currently known, including all scoring rules (e.g., plurality, Borda count, and approval voting), multi-winner plurality, supermajority, STAR-Voting, instant-runoff voting, and proportional representation schemes including D'Hondt \cite{stark20} and Hamilton \cite{blomEtal21a,blomEtal21b}).

Consider a single assorter, $A$.
The value the assorter assigns to the votes on
ballot card $b_i$ is $A(b_i)$;
the value it assigns to the votes in CVR $c_j$
is $A(c_j)$.
Let $\bar{A}^b$ denote the average value of the assorter applied to the ballot cards and let
$\bar{A}^c$ denote the average value of the assorter applied to the list of CVRs.
Define \(v := 2\bar{A}^c - 1\), the \emph{reported assorter
margin}.
(The averages might be over all ballot cards and CVRs, or only over the cards and CVRs that contain the contest, 
depending on whether card style information is used for the random selection
\cite{stark20,glazerEtal21}.)

Suppose there is a 1:1 mapping between ballot cards and CVRs for a given contest, and denote the
number of each $N_b$.
An \emph{overstatement assorter} $B$ is an affine
transformation of the original assorter applied to the votes in the CVR, minus the
assorter applied to the votes on the corresponding
ballot card:
$$
B(b_i, c_i) := \frac{1-(A(c_i)-A(b_i))/u}{2-v/u},
$$
where $u$ is an upper bound on the value that the
assorter assigns to any ballot card.
For instance, for a plurality contest, $u=1$.
\cite{stark20} shows that $B$ also assigns a bounded
nonnegative number to each ballot, and that
$\bar{A}^b > 1/2$ iff $\bar{A}^c > 1/2$ and
$\bar{B} > 1/2$, where $\bar{B}$ is 
the average value of the overstatement assorter.
Note that $\bar{B}$ can be written
\begin{eqnarray}
\bar{B} &:=& \frac{1}{N_b} \sum_{i=1}^{N_b}
    \frac{1-(A(c_i)-A(b_i))/u}{2-v/u} \nonumber \\
    &=& \frac{1}{2-v/u} \left [
     1 - \frac{1}{u} (\bar{A}^c - \bar{A}^b) \right ].
\end{eqnarray} 
Let $\pi$ be any permutation of $\{1, \ldots, N_b\}$,
that is, any 1:1 mapping from $\{1, \ldots, N_b\}$
to itself.
The mean of a list does not depend on its order, so
$$
\bar{A}^b = \frac{1}{N_b} \sum_i A(b_i) = 
\frac{1}{N_b} \sum_i A(b_{\pi(i)})
$$
and
$$
\bar{A}^c = \frac{1}{N_b} \sum_i A(c_i) = 
\frac{1}{N_b} \sum_i A(c_{\pi(i)}).
$$
Define
$$
\bar{B}^\pi  := \frac{1}{N_b} \sum_{i=1}^{N_b} B(b_{\pi(i)}, c_i).
$$
It follows that
\begin{equation} \label{eq:fundamental}
\bar{B}^\pi = \bar{B}.
\end{equation}
Hence, if the number of ballots that might
contain the contest equals the number
of CVRs that contain the contest and
$\pi$ is \emph{any} 1:1 mapping of cards to CVRs
(even a mapping that does not purport to pair a given
card with the CVR the voting system generated for
that card), the outcome of the contest is correct if
\begin{equation} \label{eq:anything_works}
\bar{B}^\pi > 1/2
\end{equation}
for every overstatement assorter $B$ for that contest.
We will exploit this result to construct the 
card-level comparison audit using the untrusted
imprinter and untrusted card retriever.

\subsection{Mismatches between the numbers of cards and CVRs} \label{sec:mismatch}

Suppose the number $N_b$ of ballot cards that might contain a particular contest is less
than the number $N_c$ of CVRs that contain 
the contest.
Then there has been a malfunction, a procedural error (e.g., some ballots were scanned
more than once or some scans or CVRs were uploaded more than once), or the integrity of the 
paper trail was compromised.

If the integrity of the paper trail has been confirmed by a compliance audit, ruling out the last possibility, the 
reported outcome can still be checked by ignoring
the contest on any $N_c -  N_b$
of the CVRs that contain the contest, provided 
it is still the case that $\bar{A}^c > 1/2$ when those CVRs are omitted.
(For audits that do not use card style information,
omitting the contest from CVR $c_i$ amounts to
setting $A(c_i) = 1/2$; for audits that use
card style information, CVRs that do not contain
the contest are simply ignored.)

After this modification of the CVRs, the number of CVRs that contain the
contest is equal to the number of cards that might contain the contest.
Let $\pi$ be \emph{any} 1:1 mapping from those CVRs to those cards, even a mapping created by a malicious adversary.
The contest outcome is correct if
$$
\bar{B}^\pi > 1/2
$$
for every overstatement assorter $B$ involved in the SHANGRLA audit of the contest.

If the number $N_c$ of CVRs that contain the
contest is less than the upper 
bound $N_b$ on the number of ballot cards that 
contain contest $c$, auditors can create $N_b-N_c$
``phantom'' CVRs that contain the contest but no 
valid vote in the contest, as described in
\cite{banuelosStark12,stark20}; the corresponding
value of $A(c_i) = 1/2$.
These CVRs will not have IDs; if the audit
selects a phantom CVR $c_i$, the corresponding value
$A(b_i)=0$, the value least favorable to the audit.
With these phantoms, the number of CVRs that contain the contest is equal to the number of cards that might
contain the contest.
Let $\pi$ be \emph{any} 1:1 mapping between those CVRs and those ballot cards, even a mapping created by a malicious
adversary.
As before, the contest outcome is correct if
$$
\bar{B}^\pi > 1/2
$$
for every overstatement assorter $B$ involved in the contest.

\subsection{Limiting risk when imprinting and retrieval are untrustworthy} \label{sec:untrust}

We present a coupling argument that reduces the problem of conducting a card-level comparison RLA when the system that 
imprints and retrieves cards is untrusted to a
standard SHANGRLA card-level comparison RLA.

We assume henceforth that there are just as many CVRs as ballot cards containing each contest; if not,
the methods in section~\ref{sec:mismatch}
can be used to make them equal.
If the system committed to a 1:1 mapping from CVR IDs to ballot cards before the audit began (i.e., committed
to which card it would retrieve when a card with the identifier $\zeta$ of the CVR $c_i$ was requested), 
a relatively simple audit procedure could
provide an RLA: sampling CVR IDs, retrieving the corresponding card (if any), and comparing the CVRs to the MVRs would provide
a random sample of overstatement errors, even if the mapping had been chosen maliciously,
so standard card-level comparison RLA methods would work.

But a faulty or malicious card-retrieval system does not need to commit to the mapping from IDs to cards 
in advance: it can adaptively determine which card (if any) to
retrieve when the auditors ask for the card with ID $\zeta$.

We address this issue as follows.
Recall that the mean of an overstatement assorter is the same for all permutations of the CVRs, as shown
in equation~\ref{eq:fundamental}.
Thus, we first construct a 
canonical 1:1 mapping $\pi$ from CVRs to cards.
For each overstatement assorter $B$,
$\bar{B}^\pi > 1/2$ iff $\bar{A} > 1/2$.

Sampling from the overstatement assorter values $\{B(b_{\pi_i}, c_i) \}_{i=1}^{N_b}$ could 
be used to test whether 
$\bar{B}^\pi \le 1/2$ using any of the methods in
\cite{stark20,waudby-smithEtal21,stark22}.
If that hypothesis is rejected for all assorters $A$ for all contests under audit,
the audit can stop.
But because the retrieval system can pick which card to retrieve \emph{after} the card with
ID $\zeta$ has been requested, there is no way to sample from $\{B(b_{\pi_i}, c_i) \}_{i=1}^{N_b}$.
Auditors can always observe $c_i$ (the auditors have the full list of CVRs), 
but might not be able to observe $b_{\pi_i}$.
So, we define a deterministic function that couples sampling from $\{B(b_{\pi_i}, c_i) \}_{i=1}^{N_b}$ 
to sampling from a related population $\{L_i\}_{i=1}^{N_b}$
for which $0 \le L_i \le B(b_{\pi_i}, c_i)$ for all $i$.
Because $L_i \le B(b_{\pi_i}, c_i)$, it follows that
$\bar{L} := \frac{1}{N_b} \sum_i L_i \le \bar{B}^\pi = \bar{B}$.
Thus if $\bar{L} > 1/2$, also $\bar{B} > 1/2$.
If we can reject the hypothesis $\bar{L} \le 1/2$,
we can conclude that the original assertion $\bar{B} > 1/2$ is true.

Testing whether $\bar{L} \le 1/2$ can be done using any of the methods in 
\cite{stark20,waudby-smithEtal21,stark22} or other valid statistical methods for testing whether
the mean of a bounded, nonnegative population is less than a given constant.
If the hypothesis $\bar{L} \le 1/2$ is rejected for every assorter
$A$, the audit can stop: all the original assertions have been confirmed statistically.

We now construct the canonical mapping $\pi$.
\begin{itemize}
   \item For each CVR ID $\zeta$:
     \begin{itemize}
      \item Let $i$ denote the index of the CVR with ID $\zeta$
      \item If $\zeta$ appears on exactly one ballot card, let 
             $\pi(i)$ be the index of that card.
      \item If $\zeta$ appears on more than one card, 
            let $\pi(i)$ be the index of the card in that
            set that maximizes $B(b_{\pi_i}, c_i)$, 
            with ties broken arbitrarily.
     \end{itemize}
   \item If some ID does not appear on any card, there are leftover IDs and an equal number of leftover cards. 
           Let $\pi$ pair their indices arbitrarily.
\end{itemize}
Now $\pi$ is a 1:1 mapping from CVR indices to cards, and the assertion $A$ is true iff
$\bar{B}^\pi > 1/2$.

We now construct a lower bound $L_i \le B(b_{\pi_i}, c_i)$ that can be calculated using the CVR $c_i$ with ID $\zeta$
and whatever card the system retrieves (or none at all) when auditors ask for the card imprinted with ID $\zeta$.
If the system returns a card with the ID $\zeta$, then $\zeta$ was imprinted on at least one card.
If it was imprinted on \emph{exactly} one card, then the value of the overstatement assorter for that 
(card, CVR) pair is just $B(b_{\pi_i},c_i)$.
If $\zeta$ was imprinted on more than one card, then the value of the overstatement assorter for that (card, CVR) pair is not larger than $B(b_{\pi_i}, c_i)$, because $\pi$ was constructed to maximize $B(b_j, c_i)$ across all cards $\{b_j\}$ imprinted with ID $\zeta$.
Thus, if $i$ is the index of the CVR with ID $\zeta$ and the ID imprinted on the card $b$ that is returned when 
ID $\zeta$ is requested is indeed $\zeta$, we can take
$L_i := B(b, c_i) \le B(b_{\pi_i}, c_i)$.

If the system does not return any card when a card with ID $\zeta$ is requested, 
if it returns a card $b$ with no ID imprinted on it, or if it returns a card $b$ imprinted with an 
ID other than $\zeta$, the true value of $B(b_{\pi_i}, c_i)$
is at least the value it would have if $A(b_{\pi(i)})=0$,
i.e.,
\begin{equation} \label{eq:ldef}
   B(b_{\pi_i}, c_i) \ge 
    \frac{1-A(c_i)/u}{2-v/u}.
\end{equation}
Thus, if the system fails to return a card with ID $\zeta$ when requested, we can take
$L_i := \frac{1-A(c_i)/u}{2-v/u} \le B(b_{\pi_i}, c_i)$.

Auditing using the values $L_i$ for randomly selected IDs $\zeta$ allows us to audit
the original assertions conservatively.

Note that if the imprinting and retrieval do what they are supposed to do, $L_i = B(b_{\pi_i}, c_i)$ for all
$i$: there is no workload penalty for the protection
against misbehavior.
But if the system does not return a card with the ID $\zeta$ when requested, in general the audit will have
to examine more ballots than it would otherwise.
In general, the workload when the outcome is correct will also depend on $\pi$, but as before, if the imprinting and retrieval are correct, $\pi$ is the identity: the audit compares the CVR for each card to the MVR for the same card.

\section{Non(c)esuch RLAs}
With these ingredients, we can now 
give the audit procedure.

\begin{itemize}

\item The voting system commits to a set of CVRs, each with an ID $\zeta$, and commits to an ID $\zeta$
(possibly blank) on each card by printing that ID on the card.
(The IDs should be pseudo-random nonces generated using an unpredictable, undisclosed seed.)
We assume that the printed IDs are immutable during the audit.

\item A risk limit is set for each contest under audit.

\item The auditors are given a trusted upper bound on the number of cards that contain each contest
under audit, a list of reported winners for each contest under audit, and the CVR list with IDs.

\item Assorters and overstatement assorters are created for every contest under audit, 
as in SHANGRLA \cite{stark20}.

\item The auditors check whether the assorter means for the CVRs are all greater than 1/2.
If not, the reported winners did not win according to the CVRs (in the case of IRV, it is possible that a different
set of sufficient assorters would all have means greater than 1/2).
The audit fails and there is a full hand count.

\item The auditors check whether the listed CVR identifiers $\{\zeta_i \}_{i=1}^{N_b}$ are unique.
If not, the audit fails.

\item The auditors next check whether the number of CVRs that contain each contest equals the number
of ballot cards that contain that contest.
If not, CVRs are altered or phantom CVRs are created as described in section~\ref{sec:mismatch}.

\item The risk-measuring function for each assertion is selected, e.g., ALPHA using the truncated shrinkage
estimator \cite{stark22} or an estimator biased towards $u$ \cite{spertusStark22}.

\item A seed for the audit's PRNG is selected, e.g., by public dice rolls.

\item The measured risk for each assorter is set to 1; all assertions and all contests are marked `unconfirmed.'

\item While at least one assertion is marked `unconfirmed':
\begin{itemize}

    \item Select an ID $\zeta$ at random from the CVR list. Let $i$ denote the index of the CVR $c_i$ with that ID.
    \item If a card with ID $\zeta$ has been requested before (which could happen if the sample is drawn with replacement), use the card (if any) previously retrieved to conduct the calculation below.
    Otherwise, ask the system to retrieve the card with ID $\zeta$.
    \item If all validly cast cards have been retrieved, determine the correct outcome of every contest from the cards and terminate
          the audit. Otherwise:
    \begin{itemize}
     \item For every contest on $c_i$
            that is under audit, for every assertion $A$ for that contest that has not yet been confirmed, 
            calculate the corresponding value $L_i$ as described in section~\ref{sec:untrust}.
            (If the system retrieved a card with ID $\zeta$, this involves manually reading the votes from that card. If no card was retrieved or if a card with a different ID was retrieved, this only involves $c_i$.)
       \item Update the measured risk for every as-yet unconfirmed assertion using the corresponding value of $L_i$.
    \end{itemize}
      \end{itemize}
    \item All assertions for which the measured risk is less than or equal to the risk limit for the corresponding
contest are marked `confirmed.'
    \item All contests for which all assertions have been marked `confirmed' are marked `confirmed'.
    \item At any time, auditors may choose to conduct a full hand count rather than continue to sample, for instance if they judge that it would be less work. If they do, the outcome according to the hand count becomes the final outcome and the audit ends.
\item End audit: all assertions have been confirmed.
\end{itemize}

This procedure can be streamlined in ways that are now conventional, for instance, by finding an initial sample size
based on the reported assorter margins such that, if the accuracy of the CVRs is above some threshold, the audit
can terminate without examining additional cards.

The procedure can also be modified to accommodate stratified sampling, e.g., using the methods in \cite{ottoboniEtal18,stark22,spertusStark22},
and to use ``delayed stratification'' \cite{stark19b}.
Sampling can also be targeted using card style information to reduce sample sizes \cite{glazerEtal21}.

\section{Implementation considerations}

The seed used to generate the nonces in each precinct should include
entropy that even an insider cannot know, so
that the IDs (and thereby the votes) cannot be linked
to the order in which ballot cards were cast.

The system should not timestamp the CVRs or digital images (or the files that contain them), or in effect the CVRs will 
have serial numbers---undermining the privacy protection the method is intended to provide.
Since some operating systems create those timestamps automatically, there may need to be a ``scrubbing'' phase to remove or randomize the timestamps before the data are exported from the scanner.

As with any system that can put marks on ballots,
it is crucial that the imprinter cannot create, alter, or obfuscate votes.
For instance, the imprinter should only use red or green ink 
and should not be physically able to print near any vote target, e.g., should only be capable of printing near the left or right
edge of a ballot card.

To facilitate the automated retrieval of cards with particular IDs, an OCR-friendly font should be used.
Barcodes or QR codes could be used, since it is not essential to
the risk-limiting property that the identifiers be human-readable (the procedure is immune to adversarial mappings), but it seems 
preferable to use human-readable marks.

\section{Conclusion}
Card-level comparison RLAs can use untrusted hardware and software to imprint putatively unique IDs on ballot cards as the 
cards are scanned and to retrieve cards that are imprinted with
particular randomly selected IDs.
The IDs can be pseudorandom nonces to protect the anonymity of the votes.
This may be particularly useful for precinct-count optical scan (PCOS) systems because
it avoids the need to re-scan ballots or to use ``hybrid'' audits, both of which are less efficient.

Untrusted technology can be used to retrieve ballot cards with specific IDs.
This may greatly decrease the workload of CLCAs, even if the IDs are serial numbers.
The non(c)esuch method can also cope with errors in the \emph{human} retrieval of ballot cards.

The non(c)esuch method relates sampling from a population of nonnegative, bounded values of a SHANGRLA 
overstatement assorter $B$ that has mean $\bar{B}$ to sampling from a related nonnegative, bounded population 
whose mean $\bar{L}$ is not larger than $\bar{B}$.
Standard statistical tests used in RLAs \cite{stark20,waudby-smithEtal21,stark22}
can be used to test the hypothesis $\bar{L} \le 1/2$.
If that hypothesis is rejected, then the hypothesis $\bar{B} \le 1/2$ can also be rejected, allowing the audit
to conclude that $\bar{B} > 1/2$.
Conducting such tests for the appropriate set of SHANGRLA assertions for each contest under audit
results in an RLA of those contests.

If the imprinting and retrieval systems behave properly, there is no performance penalty for protecting against their
possible misbehavior: the audit sample size is the same as it would be if the imprinting and retrieval 
systems were trustworthy.
In this sense, the method is adaptive.
If the imprinting or retrieval misbehaves, in general the audit will need to examine more ballot cards than it 
otherwise would to confirm contest outcomes when the outcomes are still correct.
Whether the imprinting and retrieval behave correctly or not, the procedure conservatively limits the risk to its nominal level.

Faulty or malicious implementations of the imprinting and retrieval, or 
predictability of the IDs (e.g., knowledge of the
algorithm and seed for generating putative nonces)
may compromise the anonymity of the votes, but 
cannot cause the risk to exceed the stated risk limit.

Future work will implement such a system for demonstration, using commodity hardware.

\paragraph*{Acknowledgments}
I am grateful to Marilyn Marks for helpful comments.

\bibliography{pbsBib.bib}

\end{document}